\documentclass[pra,
	twocolumn,
	showpacs,superscriptaddress,
	floatfix]{revtex4}

\usepackage{amssymb,amsmath,bm,graphicx,amsthm}
\usepackage{color}

\newcommand{\ket}[1]{|#1\rangle}
\newcommand{\bra}[1]{\langle #1|}
\newcommand{\tr}{\mathrm{tr}}

\definecolor{darkgreen}{rgb}{0,0.5,0}

\begin{document}

\title{Simulating two- and three-dimensional frustrated quantum systems\\
with string-bond states}

\author{Alessandro Sfondrini}
\affiliation{Max-Planck-Institut f\"ur Quantenoptik, 
    Hans-Kopfermann-Str.\ 1, D-85748 Garching, Germany.}
\affiliation{Dipartimento di Fisica ``Galileo Galilei'', 
    Universit\`a  di Padova, Via Marzolo 8, 35131 Padova, Italy.}
\author{Javier Cerrillo}
\affiliation{Max-Planck-Institut f\"ur Quantenoptik, 
    Hans-Kopfermann-Str.\ 1, D-85748 Garching, Germany.}
\affiliation{Institute for Mathematical Sciences, Imperial College London,
SW7 2PG, UK.}
\affiliation{QOLS, The Blackett Laboratory, Imperial College London,
Prince Consort Rd., SW7 2BW, UK.}

\author{Norbert Schuch}
\affiliation{Max-Planck-Institut f\"ur Quantenoptik, 
    Hans-Kopfermann-Str.\ 1, D-85748 Garching, Germany.}
\author{J.\ Ignacio Cirac}
\affiliation{Max-Planck-Institut f\"ur Quantenoptik, 
    Hans-Kopfermann-Str.\ 1, D-85748 Garching, Germany.}

\begin{abstract}
Simulating frustrated quantum magnets is among the most challenging tasks
in computational physics. We apply String-Bond States, a recently
introduced ansatz which combines Tensor Networks with Monte Carlo based
methods, to the simulation of frustrated quantum systems in both two and three
dimensions. We compare our results with existing results for unfrustrated
and two-dimensional systems with open boundary conditions, and demonstrate
that the method applies equally well to the simulation of frustrated
systems with periodic boundaries in both two and three dimensions.
\end{abstract}


\maketitle

\section{Introduction}

The simulation of correlated quantum spin systems is one of the central
problems in condensed matter physics. The lack of exact solutions and the
exponentially growing Hilbert space dimension motivate the need for
numerical methods for the simulation of such systems. During the last
decades, Quantum Monte Carlo (QMC)~\cite{linden:qmc} and the Density
Matrix Renormalization Group (DMRG)
method~\cite{white:DMRG,schollwoeck:rmp} have arguably been the most
successful methods for the accurate simulation of large quantum spin
systems. Despite their huge success, both methods also have their
limitations: The DMRG method gives extremely accurate results for
one-dimensional (1D) systems, but fails to simulate 2D systems similarly
well; on the other hand, QMC can deal efficiently with 2D and 3D systems,
but fails on frustrated (fermionic) quantum systems due to the so-called
``sign problem''. As frustrated quantum systems in two and three
dimensions underlie some of the most interesting phenomena in condensed matter
physics, methods which promise to overcome the previously mentioned
limitations are of high interest.

The natural generalization of the Matrix Product State (MPS) ansatz
underlying DMRG to higher dimensional systems is given by Projected
Entangled Pair States (PEPS)~\cite{frank:2D-dmrg,sierra:2d-dmrg}.
PEPS-based algorithms have been applied successfully, e.g., to the
simulation of frustrated quantum spin systems or hardcore bosons in two
dimensions~\cite{murg:peps-alg,murg:J1J2,isacsson:peps,
vidal:iPEPS-bose-hubbard,bauer:iPEPS}.  Yet, due to the scaling of
resources the method is bound to two-dimensional systems with open
boundaries, motivating the search for different tensor-network based
algorithms~\cite{vidal:mera,gu:TERG,sandvik:TERG,jiang:tns,
gu:TERGidea,jiang:tns2,sandvik:MPSsnail,evenbly:heisenberg-kagome-MERA}.
 Recently, it has been
proposed to use Monte Carlo sampling to enhance the possibilities of
tensor network based methods, both in 1D for DMRG~\cite{sandvik:mps-mc}
and, for appropriately chosen ansatz classes, in two and higher
dimensions~\cite{schuch:sbs}, and their applicability to two-dimensional
systems has been demonstrated~\cite{schuch:sbs,mezzacapo:EPS,chan:cps}.

The String-Bond States (SBS) ansatz proposed in Ref.~\cite{schuch:sbs}
generalizes the MPS ansatz to two and higher dimensions in a way
which allows to employ Monte Carlo sampling to efficiently compute
expectation values.  While the MPS ansatz is inherently one-dimensional,
SBS generalize it to higher dimensional lattices by placing several
one-dimensional structures atop of each other, e.g., along the axes, the
diagonals, and in loops between adjacent neighbors, thus allowing for
arbitrary correlations between any group of spins without
sacrificing the advantages of the one-dimensional structure.

In this paper, we demonstrate the applicability of the SBS ansatz to the
simulations of two- and three-dimensional frustrated quantum systems. In
two dimensions, we apply it to the simulation of the frustrated
$J_1$-$J_2$ model, where we find that for open boundary condition (OBC),
SBS reproduce well both the energies and the structure of correlations
obtained using the general PEPS ansatz.  Moreover, SBS also allow us to
simulate systems with periodic boundaries (PBC) with similar accuracy, and
we find that the behavior of the low-energy regime of the system in the
transition region $J_2/J_1\approx0.6$ (changing from N\'eel to columnar
order) differs significantly for OBC and PBC.

Second, we apply the SBS ansatz to the simulation of 3D frustrated spin
systems. To benchmark the ansatz, we compare results for the 3D Ising
model with transverse field to results obtained using QMC. Then, we apply
it to the simulation of a three-dimensional frustrated quantum spin system
on up to $6\times6\times6=216$ qubits, where we observe a performance
comparable to that in two dimensions.  This demonstrates the ability of
the method to simulate frustrated quantum spin systems in both two and
three dimension and with periodic boundaries.

\section{The string-bond state ansatz}

String-bond states have been proposed as a variatonal class of states for
which expectation values of local observables can be computed efficiently
using Monte Carlo sampling.  Monte Carlo sampling allows to
compute an expectation value $\sum p(n) f(n)$ over a probability
distribution $p$ by generating a sample $\{n_1,n_2,\ldots\}$ 
drawn from $p(n)$ and averaging $f$ over this sample.
Now for any observable $O$, we can rewrite its expectation value in 
a state $\ket\psi$ as
\begin{equation}
\bra{\psi}O\ket{\psi}=
    \sum_n \langle\psi\ket{n}\bra{n}O\ket{\psi}
=\sum_n p(n) \frac{\bra{n}O\ket{\psi}}{\bra{n}\psi\rangle}
\label{eq:var-mc}
\end{equation}
where $p(n)=|\langle n\ket\psi|^2$ and $\ket n$ is an orthonormal basis.
Thus, whenever $\langle n\ket\psi$ and $\langle n| O\ket\psi$ can be computed
efficiently, $\bra\psi O\ket\psi$ can be evaluated efficiently using
Monte Carlo sampling. 

We are interested in investigating systems consisting of $N$ spins with a
Hilbert space $(\mathbb C^d)^{\otimes N}$, and we thus choose the basis
$\ket{n}=\ket{n_1,\dots,n_N}$ to be a product (i.e.\ local) basis of the
system.  In order to do efficient Monte Carlo sampling in this basis, we
need that $\langle n \ket\psi$ and $\bra n O \ket\psi$ can be
computed efficiently.  The second requirement can be reduced to computing 
a few 
overlaps $\langle \tilde n\ket\psi$ whenever 
$O=\sum D_kP_k$ with $D_k$
diagonal, $P_k$ permutations, and the set of $k$'s sufficiently small,
since then $\bra n O \ket\psi= \sum_k D_k(n) \langle n_k \ket\psi$, with
$\bra n_k=\bra n P_k$ another local basis state.
In particular, this holds for local $O$ (where local means small support,
as e.g.\ the terms in a Hamiltonian or two-point correlation functions)
and tensor products of Paulis, for instance the Jordan-Wigner
transform of fermionic hopping terms, or string order parameters.

Thus, in order to be able to apply Monte Carlo sampling, we need to find
classes of states $\ket\psi$ for which the overlap $\langle n  \ket\psi$
can be computed efficiently. We choose $\langle n\ket\psi$ to be a product
of efficiently computable functions $f_s$ (with $s=1,\dots,S$) defined on subsets 
$\mathcal N_s\subset \{1,\dots,N\}$ of spins, 
\begin{equation}
\label{eq:sbs:overlap-factorize}
\langle n\ket\psi=f_1(n_{\mathcal N_1})
    \cdots
    f_S(n_{\mathcal N_S})\ .
\end{equation}
Here, $n_{\mathcal N_s}$ contains the state of all spins in the subset
$\mathcal N_s$. Note that the subsets $\mathcal N_s$ should be overlapping
as otherwise they just describe a product state.

Our choice of the $f_s$ will be such as to 
generalize Matrix Product States (MPS) to higher dimensional systems.
An MPS of \emph{bond dimension} $D$ is given by 
\begin{equation}
\ket\psi=\sum_{n_1,\dots,n_N} 
\tr\left[M^1_{n_1}\cdots M^n_{n_N}\right]\ket{n_1,\dots,n_N}
\label{eq:mps}
\end{equation}
where $M_y^x$ are $D\times D$ matrices. In order to generalize MPS 
in the spirit of the ansatz (\ref{eq:sbs:overlap-factorize}), we choose each 
$f_s$ such that 
\begin{equation}
\label{eq:onestring}
f_s(n_{i_1},\dots,n_{i_l})=
    \tr\left[M^{s,1}_{n_{i_1}}\cdots M^{s,l}_{n_{i_l}}\right]
\end{equation}
to be a trace of matrix products. Here, $i_1,\dots,i_l$ denotes the spins
in the corresponding subset $\mathcal N_s$; note that this imposes an
ordering on these sets. Clearly, this definition includes MPS themselves,
since we can choose only one $\mathcal N_s=\{1,\dots,N\}$.

\begin{figure}[t]
\includegraphics[width=0.95\columnwidth]{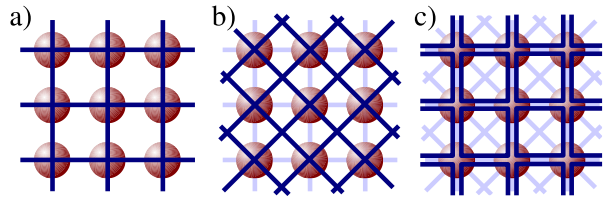}
\caption{ \label{fig:stringpatterns}
(Color online).
String patterns used in the simulations. \textbf{a)} The basic 
\emph{lines} pattern. It can be enhanced by the \textbf{b)}
\emph{diagonals} pattern and by the \textbf{c)} \emph{loops} pattern,
which help to improve the control over diagonal and four-body
correlations, respectively.
}
\end{figure}

In defining SBS on higher dimensional systems, the choice of the subsets
$\mathcal N_s$ (called ``strings'' furtheron, as they impose a 
one-dimensional ordering in the spirit of MPS) is of central importance.
The idea is that the string pattern should reflect the geometry of the
system in such a way that spins which are closely coupled by the
Hamiltonian are rather closely connected by a string. For a 2D square
lattice, a natural choice is to first put strings on all rows
(i.e., one row forms one string, corresponding to a product of MPS on
rows) and then
connect the rows by additionally placing one string per column. We call
this pattern, as illustrated in Fig.~\ref{fig:stringpatterns}a,
\emph{lines}. The \emph{lines} pattern can be enhanced in two different
ways by putting additional strings: First, one can put strings on all
diagonals of the lattice (Fig.~\ref{fig:stringpatterns}b), and second, one
can choose strings which form small loops, encompassing all elementary
plaquettes (i.e., blocks of $2\times 2$ spins,
Fig.~\ref{fig:stringpatterns}c; cf.~\cite{mezzacapo:EPS} for a
generalization of this ansatz); both of these extensions allow for a
better control of the correlations with diagonal neighbors. The patterns
generalize straightforwardly to lattices in 3D or with different
geometries. Note that by continuously adding strings, we will eventually
be able to describe all states as SBS, as can be seen by putting one long
snail-like string on the lattice (i.e., describing the whole state as an
MPS). Clearly, for good practical results, the strings should be chosen
such that the relevant states are well approximated at an early stage of
the pattern.

The computational resources of SBS scale favorable as compared to PEPS:
For each string, a matrix trace (\ref{eq:onestring}) has to be computed
which takes resources $lD^3$ ($lD^2$ for OBC), with $l\le N$ the length of the
string. This has to be multiplied by the number of strings $S$, giving a
computational cost of $O(SND^3)$. In particular, the scaling in the
accuracy parameter $D$ compares favorably to the $D^{10}$ ($D^{18}$) scaling
of the PEPS method for OBC (PBC).

Let us briefly note that although we motivated SBS as a higher-dimensional
generalization of MPS, one can also regard them as a specialized case of
PEPS. PEPS form the most natural generalization of MPS to two
dimensions~\cite{frank:2D-dmrg}, they are known to approximate the states
of interest well~\cite{hastings:locally,hastings:mps-entropy-ent}, and
have been applied successfully in numerical
simulations~\cite{murg:peps-alg,murg:J1J2}. However, the
scaling in the accuracy parameter is rather bad, preventing the
application of PEPS to problems beyond 2D systems with OBC (note, however,
that iPEPS have been applied successfully to investigate 2D systems in the
thermodynamic limit~\cite{vidal:iPEPS-bose-hubbard,bauer:iPEPS}). One way to
resolve this problem is to look for subclasses of PEPS which allow for
more efficient algorithms. Indeed, SBS form such a subclass of
PEPS~\cite{schuch:sbs}: While general PEPS are described by tensor
networks with general tensors $T_{i\alpha\beta\gamma\delta}$, SBS with a
\emph{lines} pattern have tensors of the form
$A_{i\alpha\beta}B_{i\gamma\delta}$. Note, however, that the structure of
the tensors gets more and more rich as one places additional strings on
the lattice, and thus, SBS can only be embedded in PEPS at a cost
exponential in the number of strings; moreover, since SBS computations
scale much more favorably in the accuracy parameter, even for a basic
\emph{lines} pattern SBS can outperform PEPS as they can reach much larger
$D$'s.

\section{Variational method using string-bond states}

In the previous section, we have introduced string-bond states (SBS) as a
class of states which generalize MPS to two- and higher dimensional
systems while allowing for an efficient computation of expectation values.
In this section, we will show how SBS can be used to build a variational
algorithm for simulating the ground states of quantum spin systems.
Although the ability to efficiently compute expectation values is a
necessary criterion, it is not sufficient: One also needs an efficient
and practical way to evolve the SBS towards the ground state.

The basic idea of a variational algorithm based on SBS is to fix a family
of SBS (i.e., fix a certain string pattern and the dimension $D$ of the
underlying matrices) and try to find the state within this family which
minimizes the energy of a given local Hamiltonian. Similar to DMRG or the
variational method over PEPS, we will carry out the optimization in a
local fashion: We start from some SBS, described by a number of
three-index tensors $M$ as in (\ref{eq:onestring}), select one of the
tensors -- let us call it $A$ -- and try to minimize the energy with
respect to this tensor while keeping the others fixed.  This procedure is
repeated for all tensors over and over until the energy
converges, i.e.\ a minimum within the family of states is reached.

To determine how to change the selected tensor $A$ such as to minimize the
energy, we use the linearity of the string-bond states in the tensor $A$
to be optimized,
\begin{equation}
E(\psi_A)=\frac{\bra{\psi_A}H\ket{\psi_A}}{\bra{\psi_A}\psi_A\rangle}=:
    \frac{ \bm{\bra{A} X\ket A}}{\bm{\bra A Y\ket A}}\ ,
\label{eq:qform}
\end{equation}
where we have explicitly denoted the dependence of the string-bond state
$\ket{\psi_A}$ on $A$.  $\bm{\bra A X \ket A}$ denotes a quadratic form in
$A$, where $\bm{\ket A}$ is the vectorized form of $A$, i.e.\
$\bm{A}_{(ijk)}=A_{ij}^{k}$, and we use boldface to avoid confusion with
vectors in state space.  Minimizing (\ref{eq:qform}) with respect to $A$
is a generalized eigenvalue problem and can be solved efficiently.

In order to sample $\bm X$ and $\bm Y$, define vectors
$\bm{\ket{a_n}}$ and
$\bm{\ket{b_n}}$ via the linear functionals 
\begin{equation}
\bm{\langle a_n\ket A}=
\frac{\bra{n}H\ket{\psi_A}}{\bra{n}\psi_{A_0}\rangle}\ , \
\bm{\langle b_n\ket A}=
\frac{\bra{n}\psi_A\rangle}{\bra{n}\psi_{A_0}\rangle}\ .
\label{eq:an-bn}
\end{equation}
where $A_0$ is the initial value of the tensor $A$.
It follows that the 
matrices $\bm X$ and $\bm Y$ in (\ref{eq:qform}) can be expressed
as
\begin{equation}
\bm X=\sum_n p_0(n) \bm{\ket{b_n}\bra{a_n}}\ , \ \ 
\bm Y=\sum_n p_0(n) \bm{\ket{b_n}\bra{b_n}}\ ,
\label{eq:sampleXY}
\end{equation}
where $p_0(n)\propto|\langle n\ket{\psi_{A_0}}|^2$, 
and thus determined by Monte
Carlo sampling of $\bm{\ket{b_n}\bra{a_n}}$ and   
$\bm{\ket{b_n}\bra{b_n}}$, respectively. Note that by virtue of this
definition, we obtain the normalization $\bra{\bm A_0}\bm Y\ket{\bm A_0}=1$.

However, there is a major problem with the approach of solving the
generalized eigenvalue problem: Monte Carlo sampling
$\bm X$ and $\bm Y$ is
relatively inaccurate as compared to e.g.\ the approximate contraction as
done in the PEPS algorithm~\cite{frank:2D-dmrg}, and moreover, our
estimates of $\bm X$ and $\bm Y$ get less and less accurate for $A$'s far
away from $A_0$ as we have sampled with respect to the distribution at
$A=A_0$. Specifically, already small errors in $\bm Y$ might lead to
completely wrong minima for the generalized eigenvalue problem. While for
the PEPS algorithm, this problem can be successfully overcome by
truncating small eigenvalues of $\bm Y$, this is impractical for a method
based on Monte Carlo due to the comparatively large error.

To overcome this problem, we do not solve the generalized eigenvalue
problem to compute the new $A$, but rather compute the gradient of the
energy with respect to $A$ and change $A$ slightly along this gradient 
such as to decrease the energy. First, this accounts for the fact that our
sample of $\bm X$ and $\bm Y$, Eq.~(\ref{eq:sampleXY}), is most accurate
around $A_0$, and as we will see,
 it moreover yields a formula where neither $\bm X$ nor $\bm Y$
appear in the denominator, such that small absolute errors remain small.
Another advantage will be that it is possible to gain a
considerable speed-up when sampling all the gradients simultaneously and
change all the tensors along their gradient simultaneously -- this is
possible since the gradients decouple to first order.

In order to determine the gradient of the energy with respect to $A$
around $A_0$, consider a small variation $A=A_0+\epsilon B$ (with
$\epsilon\ll 1$):
\begin{align*}
E(\psi_{{A_0}+\epsilon B})
    &=\frac{{\bra{\bm {A_0}+\epsilon\bm B} \bm X\ket{\bm {A_0}+\epsilon\bm B}}}{
	{\bra{\bm {A_0}+\epsilon\bm B}\bm Y\ket{\bm {A_0}+\epsilon\bm B}}} \\
    &=\frac{\bra{\bm {A_0}}\bm X\ket{\bm {A_0}}
		+2\epsilon\,\mathrm{Re}\left[\bra{\bm {B}}\bm X\ket{\bm A_0}\right]
		+O(\epsilon^2)}{
	1+2\epsilon\,\mathrm{Re}\left[\bra{\bm {B}}\bm Y\ket{\bm A_0}\right]
		+O(\epsilon^2)} \\
    &=E(\psi_{A_0})
		+2\epsilon\,\mathrm{Re}\left[\bra{\bm {B}}\bm X\ket{\bm A_0}\right]\\
	&\qquad-2\epsilon \bra{\bm {A_0}}\bm X\ket{\bm {A_0}}
		\mathrm{Re}\left[\bra{\bm {B}}\bm Y\ket{\bm A_0}\right]
		+O(\epsilon^2)
\end{align*}
where we have used the normalization $\bra{\bm A_0}\bm Y\ket{\bm A_0}=1$.
Thus, the gradient turns out to be
\begin{equation}
\label{eq:gradient-XY}
\nabla_A E(\psi_A)\big|_{A=A_0}=
		2\big[\bm X\ket{\bm A_0}
    -E(\psi_{A_0}) \bm Y\ket{\bm A_0}\big]
\end{equation}
[using $\bra{\bm A_0}\bm X\ket{\bm A_0}=E(\psi_{A_0})$].
Substituting the sampling formulas (\ref{eq:sampleXY}) for $\bm X$ and
$\bm Y$ and using that 
$\bm{\langle b_n\ket{A_0}}=1$ [Eq.~(\ref{eq:an-bn})], we finally obtain
that
\[
\nabla_A E(\psi_A)\big|_{A=A_0}=
2\sum_n p_0(n)\bm{\ket{b_n}}\left[
    E_n-E(\psi_{A_0})\right]\ ,
\]
where we have defined
\[
E_n:=\bm{\langle a_n\ket{A_0}}=
    \frac{\bra n H\ket{\psi_{A_0}}}{\bra n\psi_{A_0}\rangle}\ ,
\]
and the energy can be computed as
\[
E(\psi_{A_0})=
    \sum_n p_0(n) 
	\underbrace{\langle {\bm A_0}\ket{\bm b_n}}_{=1}
	    \bra{\bm a_n}{\bm A_0}\rangle = \sum_n p_0(n) E_n\ .
\]

While the previous derivation holds for any ansatz where $\ket\psi_A$ is
linear in $A$, there are some additional tricks
which can be applied in the case of SBS to save computation time.  To this
end, note that all we have to know are $\ket{\bm b_n}$, $E_n$, and the
ratio $p_0(n)/p_0(m)$ (this is sufficient to generate a random walk). For a
particular tensor $A^s_{ij}$, the dependence of $\langle n\ket{\psi_A}$ on
$A$ (where $n_A$ denotes the state of the spin associated with $A$)  can be expressed as 
\[
\langle n\ket{\psi_A}=\tr[A^{n_A}X(n)]c(n)
\]
where $X(n)$ is the product of all other matrices on the string containing
$A$ as a function of the state $n$ of the spins, and $c(n)$ contains the
contributions from all other strings. Thus, we have that
\[
\bm{\langle b_n\ket A}
=\frac{\bra{n}\psi_A\rangle}{\bra{n}\psi_{A_0}\rangle}
=\frac{\tr[A^{n_A}X(n)]c(n)}{\tr[A_0^{n_A}X(n)]c(n)}
\]
and therefore (with $s$ the physical spin index)
\[
(\bm b_n)^s=\frac{\delta_{s,n_A} X(n)^\dagger}{\tr[A_0^{n_A}X(n)]}\ .
\]
This means that in order to compute $\ket{\bm b_n}$ for a given tensor $A$, 
one only has to consider the string which contains $A$, instead of having
to look at all the strings.

Similarly, in order to compute $E_n$, one can
exploit that for local Hamiltonians, string-order operators, etc., 
$\bra n H=\sum_{m\in\mathcal M}f(m)\bra m$ where $\mathcal M$ has only 
few elements, and e.g.\ for local Hamiltonians on a 2D lattice,
each $m\in\mathcal M$ only differs at two adjacent sites
from $n$.
Thus,
\[
E_n=\frac{\bra n H\ket{\psi_{A_0}}}{\bra n\psi_{A_0}\rangle}
    =\sum_{m\in\mathcal M}
	\frac{\langle m\ket{\psi_{A_0}}}{\bra n\psi_{A_0}\rangle}
\]
can again be computed as the ratio of the matrix product traces for only
the two strings on which $m$ and $n$ differ, again reducing the
computational effort.  Computing $p_0(m)/p_0(n)$ also allows for
optimizations, depending on the way the new configuration $m$ is
constructed starting from $n$. For the simplest scenario where only a
single spin is flipped, again only the strings containing this very spin
have to be considered, and similarly if e.g.\ a pair of spins is being
flipped.

Finally, we gain a speed-up by computing the gradients for \emph{all}
tensors simultaneously; this is reasonable since the joint gradient of all
tensors is nothing but the direct sum of the individual gradients, thus,
changing all the tensor in direction opposite to the gradient by a small
amount will decrease the energy to leading order.  In this case,
computation time is saved by the fact that the same sample drawn from
$p_0(n)$ can be used, and that $E_n$ has to be computed only once.

The full algorithms looks as follows: Fix a string pattern and
corresponding bond dimensions, and choose initial configurations for all
tensors. Then, iterate the following: 1) Compute the energy and its
gradient with respect to all tensors. 2) Change all the tensors by some
small amount in the direction given by the gradient. 3) Start over at
1) with the modified tensors. Iterate this until the change in energy
becomes smaller than some threshold and declare convergence. In order to
ensure that the step along the gradient is small enough, it is advisable
to normalize the gradients such that the step remains small even for steep
gradients.

Instead of declaring convergence of the algorithm when the energy does not
change any more, one can try to increase the precision and see whether
this leads to a further improvement in energy, and only declare
convergence if it doesn't. There are three
possiblities to do so: First, one can increase the length of the Monte
Carlo sample used for computing the energy and the gradient, second, one
can try to decrease the stepwidth used to update the tensors along the
gradient, and finally, one can try to extend the variational family of
states either by increasing the bond dimension or by adding extra strings.
In all cases, it is advisable to use the previously obtained optimum as
the initial state.

\section{Numerical results}

In the following, we present numerical results obtained for two- and
three-dimensional frustrated spin systems using string-bond states.
In all cases, the Monte Carlo sampling was carried out using
single spin flip, or adjacent spin swap, Metropolis updates. The
autocorrelation time was at most 100 updates (for the structure factor of
the $J_1$-$J_2$ model in the frustrated regime), and considerably less for
local observables or non-frustrated models, even in 3D. This allowed us to
choose the Monte Carlo samples 
sufficiently long such that in all cases, the error bars were below
what could be illustrated in the plots (local observables to at least
0.1\%, and non-local observables to at least 1\% accuracy). Note, however,
that this only means that we have good control over the error we make in
measuring observables on the given variational state; the major (and not
so well controlled) error source in the method is thus the ability of the
ansatz class to correctly describe the ground state, together with the
question as to whether the variational method converges to the optimal
state within the class.

\begin{figure}[t]
\includegraphics[width=0.9\columnwidth]{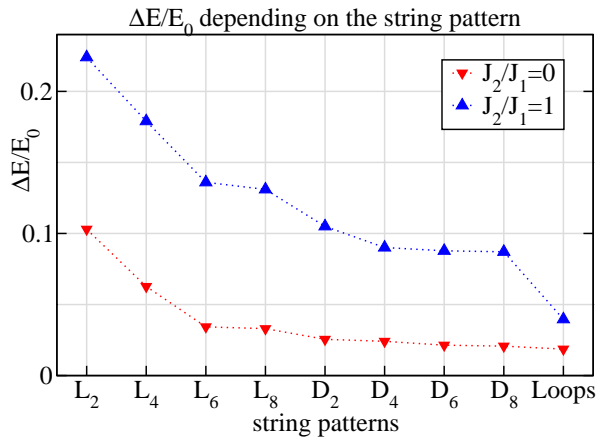}
\caption{\label{fig:improvement}
(Color online.)
Relative error in ground state energy as a function of the string pattern
for a the $J_1$-$J_2$ model on a $6\times6$ OBC lattice.
From left to right: \emph{lines} with $D=2,4,6,8$ (labelled $L_2,
\dots, L_8$), \emph{lines} with $D=8$
together with \emph{diagonals} with $D=2,4,6,8$
(labelled $D_2,\dots,D_8$), and finally \emph{lines}
with $D=8$, \emph{diagonals} with $D=8$, and loops ($D=4$). The lower
(red) points are for the Heisenberg model, $J_2/J_1=0$, and the upper
(red) blue for $J_2/J_1=1$. In the highly frustrated regime, adding
diagonals and loops leads to a significant improvement.
Note that an extrapolation is difficult to perform, as it is unclear how
the accuracy will scale in the string pattern.
}
\end{figure}

\subsection{Simulation in 2D: The $J_1-J_2$ model}

We have applied SBS to the simulation of the so-called $J_1$-$J_2$ model,
\[
H_{J1J2}=\sum_{<i,j>}{\bm\sigma}_i\cdot{\bm\sigma}_j + \frac{J_2}{J_1}
    \sum_{\ll i,j\gg}{\bm\sigma}_i\cdot{\bm\sigma}_j 
\]
where $<i,j>$ denotes nearest neighbors in a 2D square lattice, and $\ll
i,j \gg$ nearest neighbors along the diagonal. This model arises e.g.\ in
the context of the Hubbard model which is believed to underly
high-temperature superconductivity~\cite{inui:j1j2-hubb}, and has become
one of the paradigmatic models to understand quantum phase transitions in
frustrated spin systems~\cite{schulz:j1j2-6x6-ex}.

\begin{figure}[t]
\includegraphics[width=0.75\columnwidth]{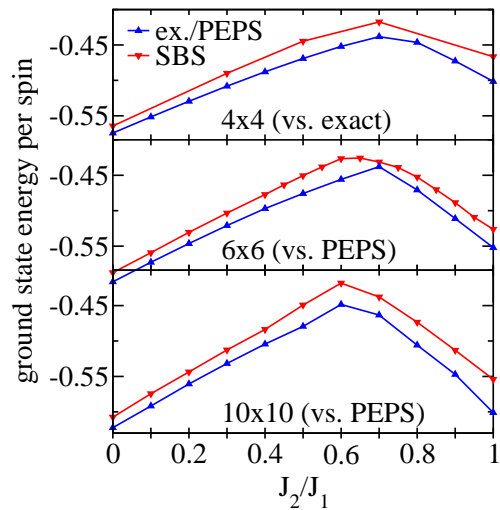}
\caption{ \label{fig:j1j2-en-obc}
(Color online).
Energy comparison for the $J_1$-$J_2$ model for OBC lattices of size
$4\times 4$ (compared to the exact energies), $6\times 6$, and $10\times
10$ (both compared to the PEPS energies~\cite{murg:J1J2}).
}
\end{figure}

\begin{figure}[h]
\includegraphics[width=0.75\columnwidth]{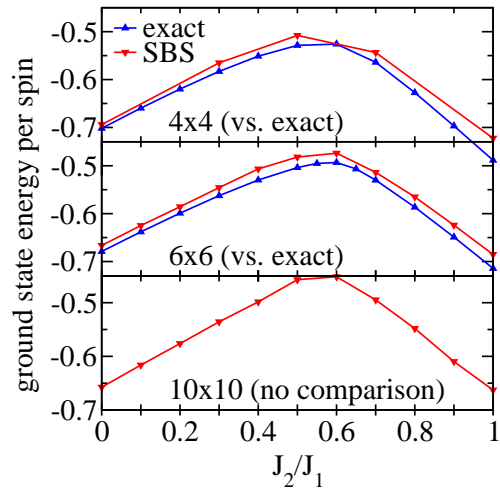}
\caption{ \label{fig:j1j2-en-pbc}
(Color online.)
Energy comparison for the $J_1$-$J_2$ model for PBC lattices of size
$4\times 4$, $6\times 6$ (both compared to exact
energies~\cite{schulz:j1j2-6x6-ex}), and $10\times
10$ (where there is no data to compare with).
}
\end{figure}

\begin{figure*}[t]
\includegraphics[width=0.95\textwidth]{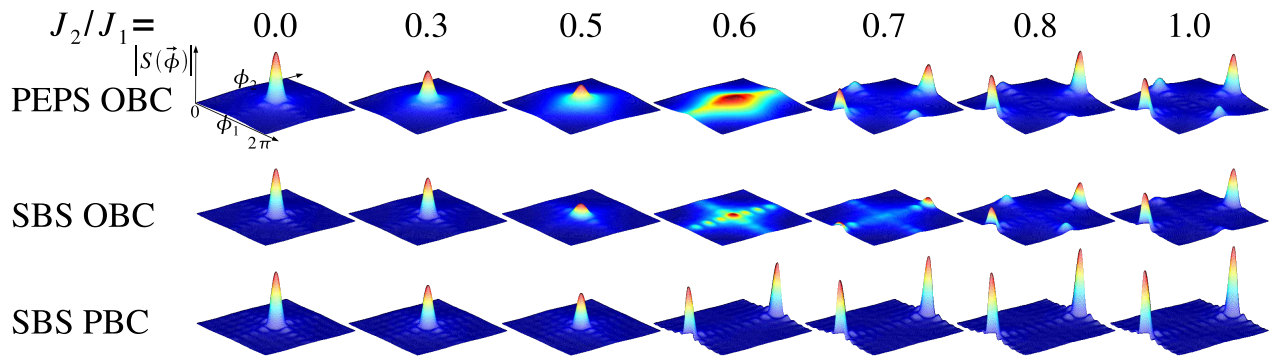}
\caption{\label{fig:j1j2-sf}
(Color online).
The absolute value of the structure factor $S(\phi_x,\phi_y)$ as defined
in Eq.~(\ref{eq:num:sfact}) computed for the $J_1$-$J_2$ model on a
$10\times 10$ lattice as a function of the ratio $J_2/J_1$. The plot
compares the results obtained on an OBC lattice using
PEPS~\cite{murg:J1J2} with both the OBC and the PBC result found using
SBS. One finds that for OBC, SBS reproduce the characteristics of the PEPS
results, and there is the signature of a intermediate glassy phase around
$J_2/J_1=0.6$. For PBC, on the contrary, there is no signature of an
intermediate phase, which is missing for PBC. Note that this observation
should be taken with care,
 as the SBS energies are typically a few percent above
the PEPS. The wave-like artifacts which can be seen especially
for OBC around $J_2/J_1=0.6$ are probably due to the fact that the 
string pattern has preferred axes.
} 
\end{figure*}

For the simulation, we started from the patterns \emph{lines}, then added
\emph{diagonals}, and finally \emph{loops}.  Fig.~\ref{fig:improvement}
shows how the energy improves as $D$ is increased and additional strings
are added, for $J_2/J_1=1$;  as on one can see, the improvement due to
additional strings depends on the model under consideration.  Note that
for our simulations, we have used the $\mathrm{SU}(2)$ invariance of the
model, which implies that  we can project our ansatz into the spin $0$
subspace (as there is a ground state with spin $0$). This can be
understood as an SBS with one additional string which covers the whole
lattice and enforces $S_z=0$. In practice, we achieve the restriction by
sampling from the $S_z=0$ subspace: we start from a configuration in this
subspace and create new configurations by swapping a randomly chosen pair
of spins; we have observed that this restriction led to a significant
improvement in energy.

In Fig.~\ref{fig:j1j2-en-obc}, we show results for the ground-state energy
of the $J_1$-$J_2$ model on lattices of size $4\times 4$, $6\times 6$, and
$10\times 10$ with open boundaries, which we compare with the values
obtained using exact diagonalization ($4\times
4$) and the PEPS method~\cite{murg:J1J2} ($6\times 6$, $10\times 10$).
Fig.~\ref{fig:j1j2-en-pbc} shows the same numbers for the case of periodic
boundaries, compared to the exact numbers~\cite{schulz:j1j2-6x6-ex}
($4\times 4$, $6\times 6$). Let
us note that for $10\times 10$ lattices, there are no numbers available to
compare with.
Typical $D$'s were
$D_\mathrm{line}=D_\mathrm{diag}=6$ for up to $6\times 6$ and $8$ to $10$
for $10\times 10$, and $D_\mathrm{loop}=4$.

\begin{figure}[b]
\includegraphics[width=0.85\columnwidth]{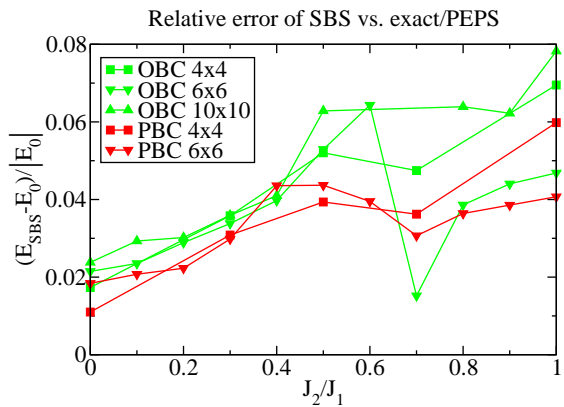}
\caption{ \label{fig:j1j2-en-relerr}
(Color online).
Relative error for the comparisons in Figures~\ref{fig:j1j2-en-obc}
and~\ref{fig:j1j2-en-pbc}. As one can see, the accuracy stays constant
when increasing the lattice size.
}
\end{figure}

The relative errors in energy corresponding to Figs.~\ref{fig:j1j2-en-obc}
and~\ref{fig:j1j2-en-pbc} are shown in Fig.~\ref{fig:j1j2-en-relerr}.
While the energies obtained using SBS are above the exact/PEPS data, 
the error does not seem to depend on the system size or the choice of
boundaries, which suggests that the method should be equally applicable to
larger and PBC systems.

Let us now see whether SBS can reproduce the correlation functions of the
$J_1$-$J_2$ model.  To this end, we use the structure factor
\begin{equation}
\label{eq:num:sfact}
S(\vec\phi)=\sum_{\vec n,\vec m}e^{i (\vec n-\vec m)\cdot \vec\phi}
    \langle {\bm\sigma}_{\vec n}\cdot{\bm\sigma}_{\vec m}\rangle\ .
\end{equation}
$S(\vec\phi)$ is the Fourier transform of the two-point correlation
functions $\langle {\bm\sigma}_{\vec n}\cdot{\bm\sigma}_{\vec m}\rangle$,
i.e., it reveals information about the relative alignment of the spins,
this is, the order of the system~\cite{auerbach:book}.
  The results for for PEPS with OBC, SBS with OBC, and
SBS with PBC is diplayed in Fig.~\ref{fig:j1j2-sf}. Note that the OBC
results exhibit the same characteristic properties for both PEPS and SBS,
while the SBS results for PBC are significantly different in the
region around $J_2/J_1=0.6$, where the behavior of the model is not yet
fully understood. It is believed that in this region, the system is in
some kind of glassy phase. While a signature of this phase can be seen in
the case of OBC both for the PEPS and the SBS data, the same signature is
completely absent in the case of periodic boundaries.  While this seems to
suggest that the behavior of the system in that region might be different
for OBC and PBC, we would like to stress that this is obtained from
configurations with energies clearly above the exact and PEPS data, and
thus should be treated with care.

\begin{figure}[t]
\includegraphics[width=0.9\columnwidth]{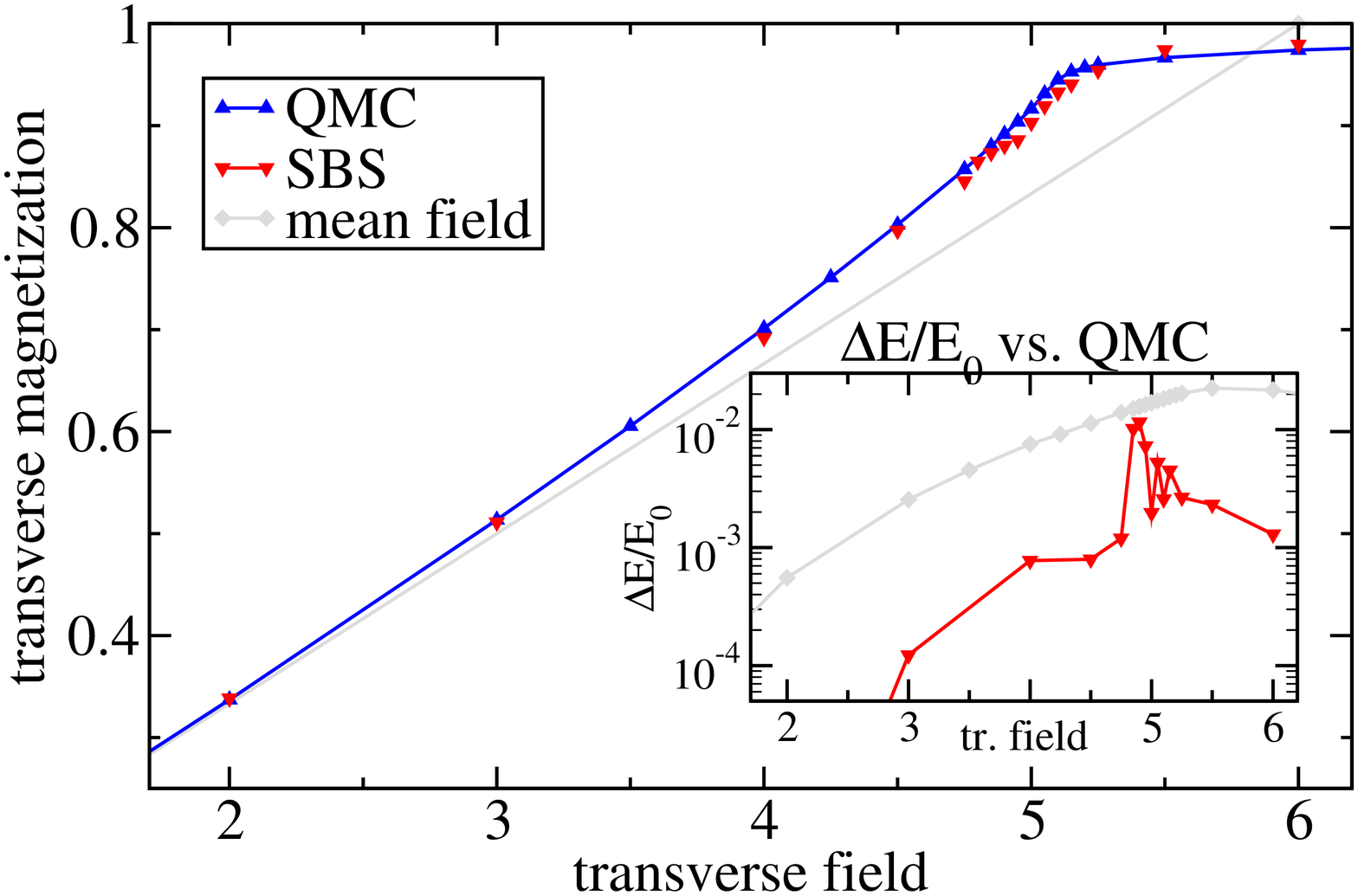}
\caption{\label{fig:3d-ising}
(Color online).
Results for the 3D Ising model in transverse field on an $8\times 8\times
8$ PBC lattice. The plot shows the magnetization obtained with SBS using the
\emph{lines} pattern, compared to QMC data obtained using
ALPS~\cite{alps:looper1,alps:looper2}, and to the mean field solution. 
The inset shows the relative
error in the energy.
}
\end{figure}

\subsection{Three-dimensional systems}

While some variational methods based on tensor networks such as
PEPS~\cite{murg:J1J2}, MERA~\cite{evenbly:heisenberg-kagome-MERA}, or Monte-Carlo
based ansatzes such as SBS~\cite{schuch:sbs} or EPS~\cite{mezzacapo:EPS}
have been shown to be able to simulate two-dimensional frustrated quantum
systems, none of the previous methods has yet be applied to the
simulation of systems in three dimensions. In the following, we give
results of SBS simulations for three-dimensional frustrated quantum
systems which are comparable to those obtained in two dimensions.

\begin{figure}[t]
\includegraphics[width=0.9\columnwidth]{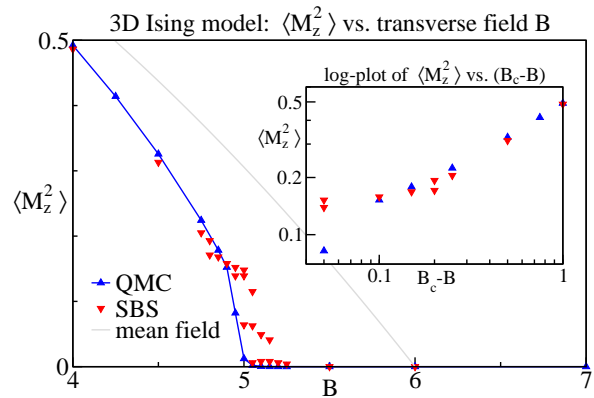}
\caption{\label{fig:3d-ising-critexp}
(Color online).
Squared magnetization $\langle M^2_z\rangle$ for the 3D transverse Ising
model on an $8\times 8\times 8$ PBC lattice, comparing data obtained with
SBS using the \emph{lines} pattern, QMC data obtained using
ALPS~\cite{alps:looper1,alps:looper2}, and the mean field solution.  The
inset shows a log-log-plot close to the critical point.  See text for a
comment of the fluctuations which can be observed around the critical
point.
}
\end{figure}

The 3D simulations are based on the \emph{lines} pattern on a 3D lattice
with PBC, and $D=6$.  To benchmark the method, we have simulated the 3D
Ising model with transverse field, $H=\sum Z_iZ_j+B\sum X_i$, on an
$8\times 8\times8$ PBC lattice, and compared the result to QMC simulations
carried out using the ALPS package~\cite{alps:looper1,alps:looper2}, as
well as mean field data.  Fig.~\ref{fig:3d-ising} shows the magnetization
along $x$ and the relative error in energy (inset) as a function of the
field $B$, and Fig.~\ref{fig:3d-ising-critexp} the magnetization squared
along the Ising coupling, $\langle M^2_z\rangle=\sum_{ij}\langle
Z_iZ_j\rangle/N^2$.  Note that the method becomes unstable close to
the critical point and frequently gives too large values for $\langle
M^2_z\rangle$. 
This is not a problem of the Monte Carlo sampling, which yields
$\langle M_z^2\rangle$ with an accuracy of about 1\% (with $1.6\times10^6$
Metropolis updates, where $\langle M^2_z\rangle$ is sampled on every
100th configuration).  Rather, this effect is due to the fact that
variational methods using MPS and related ansatzed such as SBS generally
tend to break symmetry close to the critical point even in 1D, as has been
also observed elsewhere~\cite{sandvik:sym-breaking}. This can be
understood in two ways: Firstly, the entanglement entropy of the ground
state diverges at the
critical point, so that the ground state cannot be exactly reproduced by
states such as MPS or SBS which obey an area law, thus driving the ansatz
into symmetry-broken solutions with
slightly higher energy but less entanglement.  Secondly, variational
ansatzes have a general tendency to break symmetries as this corresponds
to having less (connected) long-range correlations, and establishing such
correlations is difficult to accomplish by doing local optimizations.
E.g., in the most extreme case, once the matrices in the MPS or SBS do not
have full rank any more, the subspace not used by the matrix is lost for
the optimization as it cannot be seen any more by local variations, and in
particular by a gradient search.

\begin{figure}[b]
\includegraphics[width=0.7\columnwidth]{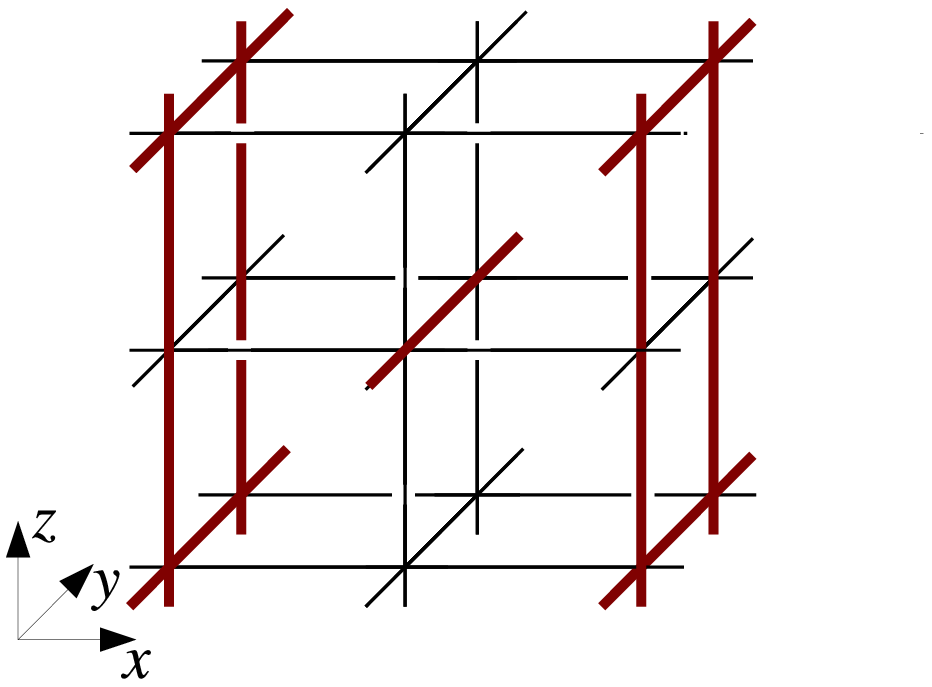}
\caption{\label{fig:3d-frustpattern}
(Color online).
Coupling pattern for the 3D frustrated XX model Eq.~(\ref{eq:num:frXX}),
illustrated for an elementary cell of size $2\times1\times2$. The thick
red edges represent ferromagnetic couplings $J_{ij}=-1$, while the other
edges correspond to antiferromagnetic couplings, $J_{ij}=1$. Note that the
role of ferromagnetic and antiferromagnetic couplings, as well as the axes
of the model, can be swapped by local $\sigma_z$ transformations.
}
\end{figure}

\begin{figure}[t]
\includegraphics[width=0.9\columnwidth]{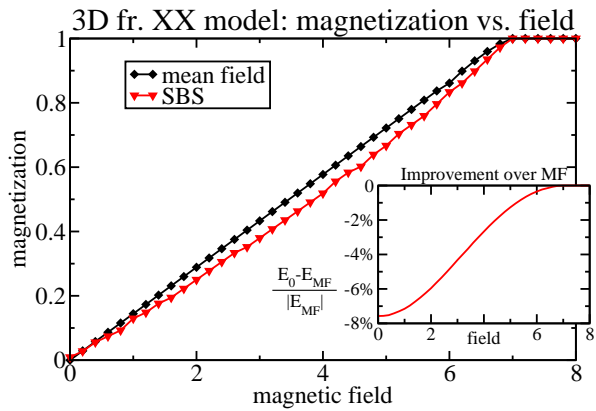}
\caption{\label{fig:666}
(Color online).
Magnetization of the 3D frustrated model (\ref{eq:num:frXX}) as a
function of the external field for a $6\times6\times6$ lattice, for SBS
and mean field.  The inset shows the improvement in energy of SBS
relative to mean field. }
\end{figure}

\begin{figure}[b]
\includegraphics[width=0.9\columnwidth]{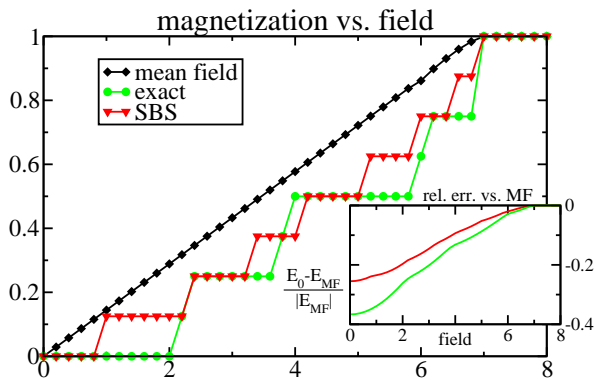}
\caption{\label{fig:224}
(Color online).
Benchmark for the 3D frustrated model (\ref{eq:num:frXX}) on a
$2\times2\times 4$ lattice: We compare the exact values with data obtained
using mean field and SBS.
The figure shows the magnetization as a function of the magnetic field,
and the inset the improvement in energy relative to mean field.
}
\end{figure}

After having tested our 3D algorithm on the Ising model, we have
subsequently applied SBS to simulate a frustrated XX model in a transverse
field on a 3D square lattice,
\begin{equation}
\label{eq:num:frXX}
H=\sum_{<\!i,j\!>}J_{ij}\left[\sigma^x_i\sigma^x_j+
    \sigma^y_i\sigma^y_j\right] 
    + B \sum_i \sigma^z_i\ ,
\end{equation}
where $<\!\!i,j\!\!>$ denotes nearest neighbors on the 3D square, and
$J_{ij}=\pm 1$ is chosen such that the system is frustrated around every
plaquette, as illustrated in Fig.~\ref{fig:3d-frustpattern}. There are
several reasons for chooosing this model: First, it is frustrated 
and thus cannot be simulated by QMC due to the sign problem.  Second, its
lower symmetry as compared to an $\mathrm{SU(2)}$ invariant model makes it
easier to simulate. Finally, for this model, the $z$ magnetization $S_z$
is a good quantum number.  Thus, the behavior of the model can be
completely understood if the minimal energy $E_m$ for fixed $m\equiv S_z$
at zero field is known: The minimal energy within each subspace with given
magnetization $m$ decreases linearly with the field, $E_m(B)=E_m-Bm$, and
the magnetization $m$ at a given $B$ is the one for which $E_m(B)$ becomes
minimal.

We have computed the $E_m$ for the model (\ref{eq:num:frXX}) on a
$6\times6\times6$ lattice and from this data determined the ground state
energy and the magnetization as a function of the field. The results are
shown in Fig.~\ref{fig:666}, where we compare it to mean field data, which
we also used to bootstrap the SBS ansatz. We found that most of the
improvement is already obtained for $D=2$ ($D=1$ being mean field), and
for $D=6$, the method was fully converged.

In order to estimate the performance of the ansatz, we have compared both
mean field and SBS to the exact solution on a $2\times2\times4$ lattice.
 The results are shown in Fig.~\ref{fig:224}: While both energy
and magnetization are still away from the exact solution, the values
obtained using SBS are significantly more accurate than the mean field solution.

\section{Conclusions}

In this work, we have presented numerical results obtained with the
recently introduced String-Bond State (SBS) ansatz for frustrated quantum
spin systems in both two and three dimensions and for open and periodic
boundaries. While the results obtained for 2D OBC systems were above the
results found using PEPS, the more favorable scaling of the method allowed
us to go beyond 2D and OBC and obtain similarly accurate results for 2D
PBC and 3D frustrated systems, which often cannot be simulated otherwise.

The computational resources needed for the simulation are moderate, as the
contraction of the strings scales only with $D^3$, and the $D$'s used are
much smaller than those in DMRG; typical simulations for the $J_1$-$J_2$
model took less than two days using a MATLAB code on a single processor.
The method allows for parallelization in evaluating energy and
gradient on the Monte Carlo sample with low interprocess communication.
Optimizations are possible with respect to caching contracted strings and
reusing them in consecutive Monte Carlo samples, as well as in 
reusing Monte Carlo samples after small updates.

There are two main challenges in the implementation of the algorithm.
First, one needs a systematic way of growing the string pattern which is
suitable for the problem at hand. As one can see in
Fig.~\ref{fig:improvement}, the same string patterns lead to different
improvements depending on the underlying model. Related to this, the
performance of the method on non-$\mathrm{SU}(2)$ invariant models will
also depend on the choice of the local basis in which the sampling is
performed, since this will affect the probability distribution sampled
over.  The second important point is to choose the proper initial state
for the optimization.  In particular, we have observed for the
three-dimensional frustrated XX model presented in the paper that the
algorithm performs much better when starting from the mean field solution
as compared to a random initial state. Here, it seems that the important
information is the proper sign of the wavefunction rather than the
amplitude, as the latter can be easily changed by the gradient flow.
(Note however that the performance for the $J_1$-$J_2$ model did not
depend on the choice of the initial state.) The proper choice of the sign
pattern will likely also pose a central challenge when applying SBS to
fermionic systems; note however that this might be overcome by using
fermionic SBS, analogous to fermionic
PEPS~\cite{kraus:fPEPS,vidal:fPEPS-sim}, instead of mapping the system to
spins via a Jordan-Wigner transform.

\subsection*{Acknowledgements}

We thank F.~Mezzacapo, F.~Verstraete, and M.~Wolf for helpful discussions
and comments, S.~Todo for help on the ALPS package, and V.~Murg for
providing us with the PEPS data. The Quantum Monte Carlo simulations of
the 3D Ising model have been carried out using the ALPS looper
code~\cite{alps:looper1,alps:looper2}, see http:/\!/alps.comp-phys.org
and http:/\!/wistaria.comp-phys.org/alps-looper.
This work has been supported by the EU
(QUEVADIS, SCALA), the German cluster of excellence project MAP, the
DFG-Forschergruppe 635, and the AXA Research Fund.

\end{document}